\documentstyle[preprint,aps]{revtex}
\tightenlines
\input{psfig}
\begin{document}
\preprint{UNDPDK-98-02}
\title{Problems with Atmospheric Neutrino Oscillations}
\author{J.M.~LoSecco}
\address{University of Notre Dame, Notre Dame, Indiana 46556}
\date{\today}
\maketitle
\begin{abstract}
The neutrino oscillation hypothesis does a poor job of representing the
atmospheric neutrino anomaly.  The anomaly is observed over 4 decades in
path length and at least a factor of 30 in energy.  This restricts the
potential oscillation solutions to those with large amplitudes and mixing
mass differences that are ruled out by other observations.  The
$\Delta m^{2}$ region in the range $10^{-4} - 10^{-2}$ leads to
inconsistencies within the atmospheric neutrino data itself.
The observed value of $R$ seems to be incompatible with the
$\Delta m^{2}$ implied by recent results\\
Subject headings: Cosmic Rays --- Elementary Particles --- Neutrino Oscillations
\\
\end{abstract}


The atmospheric neutrino anomaly \cite{haines,kamioka,imbo} is the discrepancy
between the observed and expected rate of electron and muon neutrino
interactions in underground detectors.  In general it is believed that
these neutrinos originate in the Earth's atmosphere as a consequence of
the decay of short lived particles created by cosmic ray interactions.

The atmospheric neutrino anomaly is characterized by a low fraction of
muon neutrino interactions relative to the observed rate of electron neutrino
interactions\cite{supcont}.  Only 61\% of the expected rate is found.
The deficiency is energy independent in the
range 200 MeV to 4000 MeV.  The deficiency appears to be isotropic up to
at least 120 MeV.
In particular neutrinos entering the detector from above
evidence the same muon neutrino deficiency as in other directions.

The Kamioka experiments like to present their results on the atmospheric
neutrino anomaly in terms of $R$.  $R$ is interpreted as the fraction of
muon neutrino events observed compared to what is expected.  $R$ is defined
as
\[
R = \frac{(\mu / e )_{DATA}}{(\mu / e )_{MC}}
\]

\begin{figure}
\psfig{figure=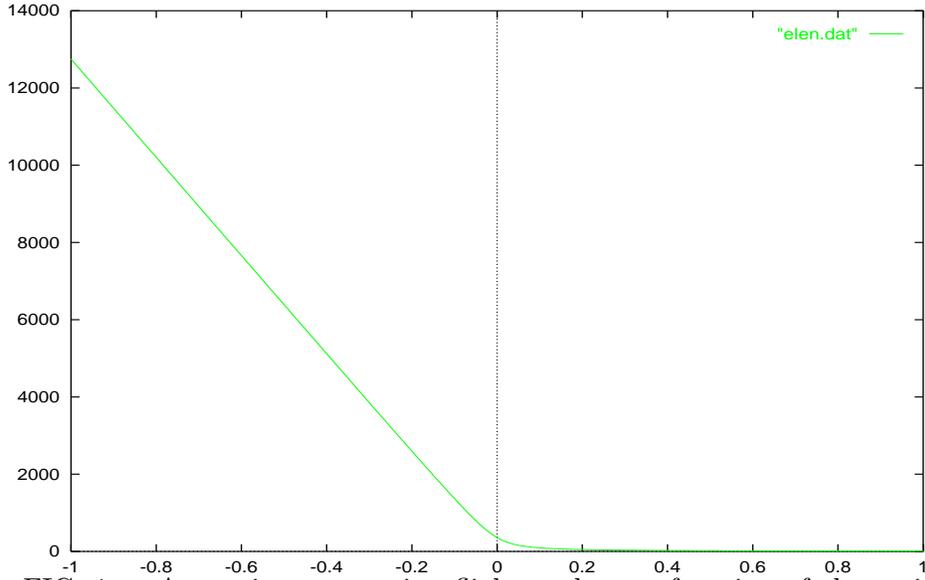,width=5.1in,height=3.0in}
\caption{\label{linear} Approximate neutrino flight path as a function of the
cosine of the zenith angle.  Distances are in kilometers.}
\end{figure}

Underground detectors sample the neutrino flux from all directions.
Since the neutrinos are produced by particle decays in the atmosphere,
different directions correspond to different neutrino flight distances.
Neutrinos coming from below the detector are primarily from a distance
of the order of 8,000 km.  Those from above have a flight path on the order
of 20 km.  Since a decay can occur anywhere in the atmosphere the decay
distance is uncertain by about 10 km.  The transition from the short flight
distance to the long flight distance is rather rapid (figure \ref{linear}).  
Most of the solid
angle is composed of sources at the two distances.  40\% of the solid angle
corresponds to a flight length under 45 km.  40\% of the solid angle
corresponds to a flight length in excess of 2600 km.  The neutrino spectrum
peaks at 200 MeV but the contained sample extends out to about 1300 MeV.

Figure \ref{linear} shows the approximate neutrino flight path as a
function of the zenith angle.  The path length as a function of zenith angle
can be represented (approximately) by
$L(\cos(\theta_{z})) = \sqrt{R_{1}^{2} (\cos^{2}(\theta_{z})-1) + R_{2}^{2}}
-R_{1} \cos(\theta_{z})$  With $R_{2}$ representing the distance from the
center of the earth to the upper atmosphere where the neutrinos are born and
$R_{1}$ representing the distance from the center of the earth to the detector
($R_{2}>R_{1}$).

The most popular explanation for the source of the anomaly is the oscillation
of muon neutrinos.
Recently the Super-Kamioka collaboration has provided strong evidence
for neutrino oscillations as a source of the atmospheric neutrino
anomaly\cite{suposc,kajita}.  But these results seem to be internally
inconsistent and inconsistent with other recent Super-Kamioka
observations of atmospheric neutrinos\cite{multig} and much of the prior
work on the anomaly\cite{haines,kamioka,imbo}.

Oscillations must be capable of
mixing the neutrino flux over the energy and distance scales observed.
The oscillation length is given by:
\[
L_{\nu} = \frac{2 \pi}{1.27 \Delta m^{2}} E_{\nu}
\]
or
\[
\Delta m^{2} = \frac{2 \pi}{1.27} E_{\nu} / L_{\nu}
\]

So the neutrino energy and oscillation distance set the scale for the
neutrino mass.  At $E_{\nu}=600$ MeV and $L_{\nu}=120$ km
(4 times 30 km since the amplitude
is proportional to $\sin^{2}(1.27 \Delta m^{2} \frac{L}{E})$)
gives a mass scale of $2.5 \times 10^{-2}$ eV$^{2}$.  At lower mass scales the
downward flux will not have had a chance to fully oscillate.  Isotropy
would no longer be apparent.

For mass scales below these it also becomes very
difficult to obtain the
observed deficiency in the global sample
since the upper hemisphere would not have
oscillated in the short distance available
and the maximum attenuation from the long path length lower
hemisphere would be 50\% (if no other oscillation length were comparable).
So the maximum attenuation of the global muon neutrino signal under these
conditions would be on the order of 75\%.

To explain the isotropic reduction of the contained event sample in the
momentum range less than about 1 GeV/c$^{2}$ requires a 
$\Delta m^{2}$ in the range of $10^{-1} - 10^{-2}$ eV$^{2}$.  But the
reduction appears to be constant out to about 4 GeV\cite{multig} which implies
a mass near the upper part of this range.
IMB has presented evidence that the isotropy
extends out to at least a mean energy of 4 GeV \cite{clark97}.
Super-Kamiokande indicates a value of $R$ is about 0.6 at 4 GeV\cite{multig}.
These observations  push the
minimum mass scale possible up to $1.7 \times 10^{-1}$ eV$^{2}$.

These arguments about isotropy and maximal flux reduction are well known
and have motivated the energy-distance scale of a number of long baseline
accelerator and reactor neutrino experiments designed to study the
oscillation hypothesis.  Such regions have been probed, and limited by studying
energetic upward going muons in underground detectors\cite{svob}.
Upward going muons come from the interactions of energetic muon neutrinos
in the rock underneath the detector.  They, in general, come from a much higher
region of the neutrino energy spectrum.

\begin{figure}
\psfig{figure=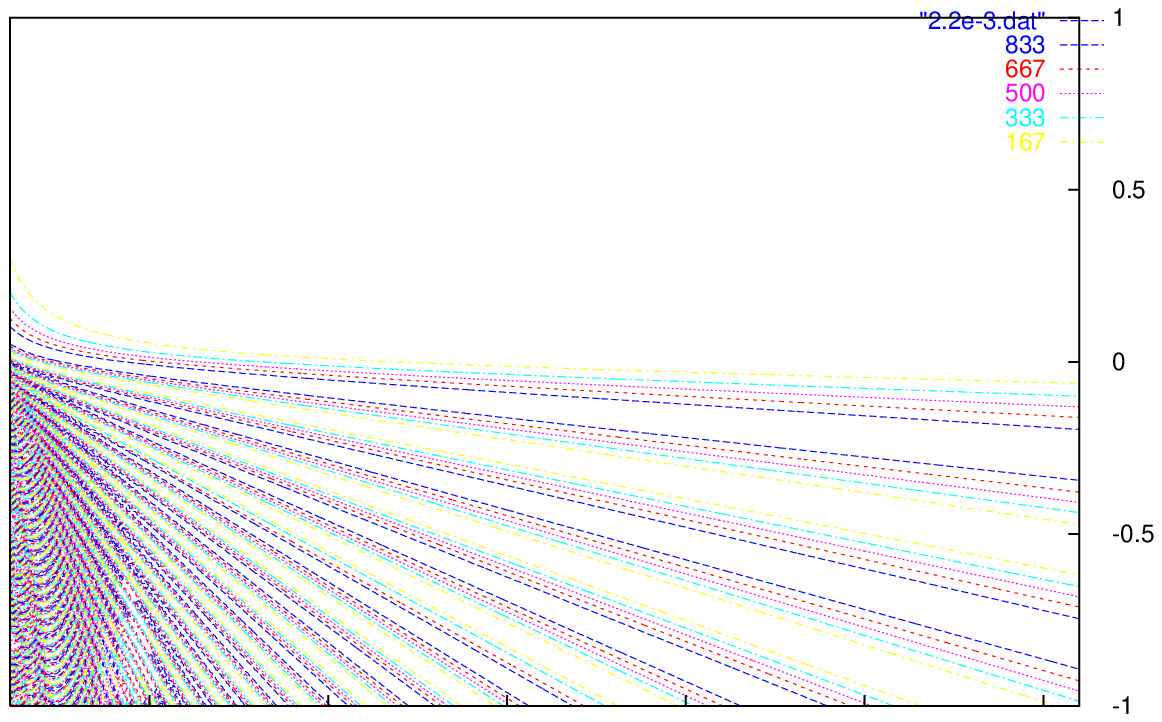,width=5.1in,height=3.0in}
\caption{\label{2.2e} Maximum attenuation of the neutrino flux as a function
of energy and cosine of the zenith angle.  A mass
scale $\Delta m^{2}=2.2 \times 10^{-3}$ eV$^{2}$ was used for this plot.}
\end{figure}

\begin{figure}
\psfig{figure=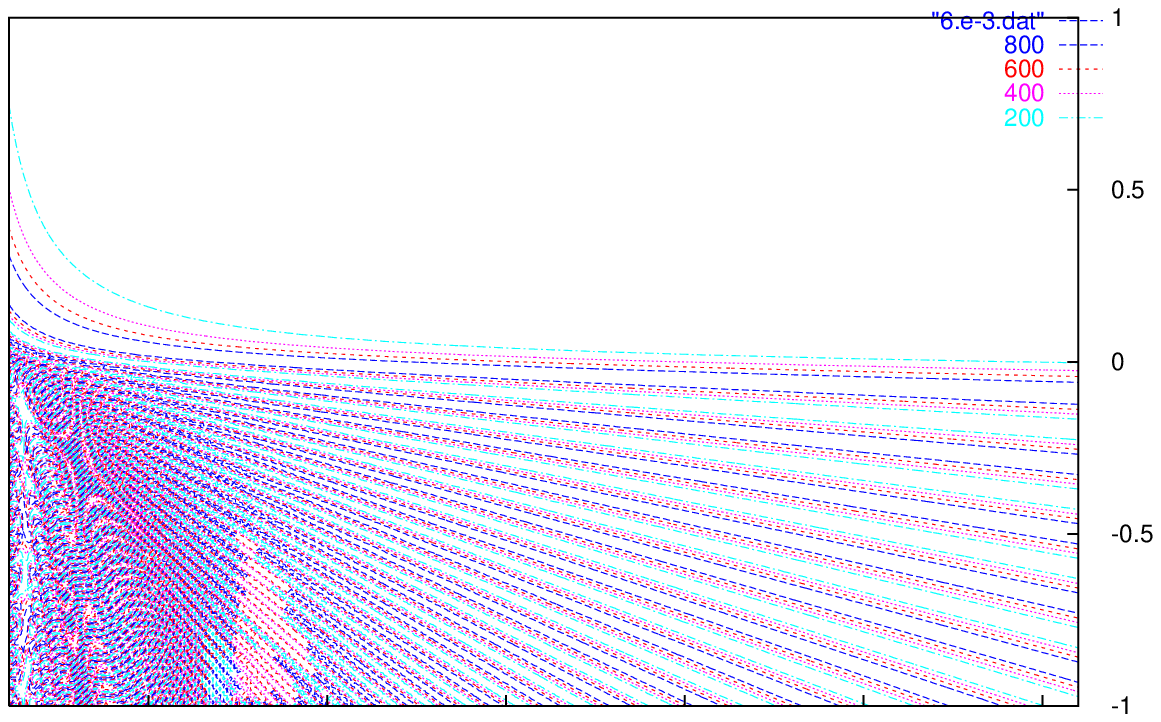,width=5.1in,height=3.0in}
\caption{\label{6e} Maximum attenuation of the neutrino flux as a function
of energy and cosine of the zenith angle.  A mass
scale $\Delta m^{2}=6. \times 10^{-3}$ eV$^{2}$ was used for this plot.}
\end{figure}

\begin{figure}
\psfig{figure=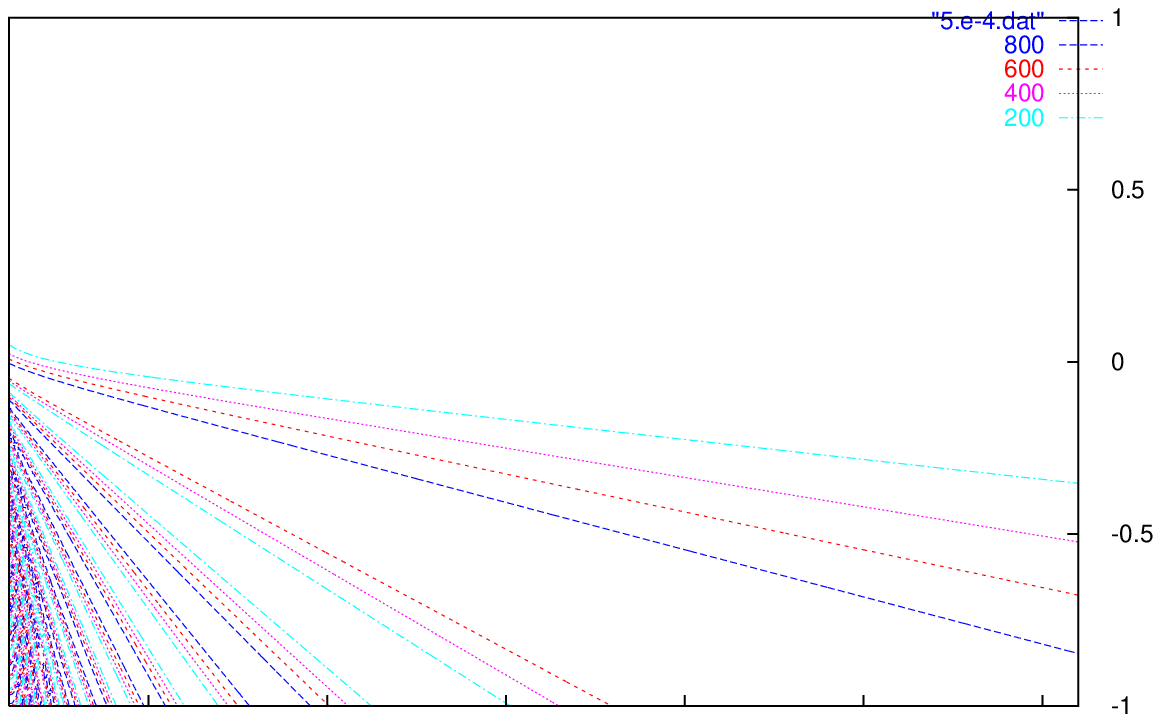,width=5.1in,height=3.0in}
\caption{\label{5e} Maximum attenuation of the neutrino flux as a function
of energy and cosine of the zenith angle.  A mass
scale $\Delta m^{2}=5. \times 10^{-4}$ eV$^{2}$ was used for this plot.}
\end{figure}

Figure \ref{2.2e} shows a contour plot of the function
$\sin^{2}(1.27 \Delta m^{2} \frac{L(\cos(\theta_{z})}{E})$ for a range of
neutrino energies from 200 MeV to 6200 MeV.  The plot is made for the mass
scale $\Delta m^{2}=2.2 \times 10^{-3}$ eV$^{2}$ favored by the recent
Super-Kamioka results.  The lowest contour present in the figure represents
a 17\% reduction in flux.  Almost no reduction is seen for neutrinos
originating in the upward hemisphere.  Figure \ref{6e} and figure \ref{5e}
are similar plots for $\Delta m^{2}=5 \times 10^{-3}$ eV$^{2}$ and
$6 \times 10^{-4}$ eV$^{2}$ respectively.

Integrals over portions of Figures \ref{2.2e}, \ref{6e} and \ref{5e} give
the maximum decrease in muon neutrino flux attributable to oscillations
for that interval.  (Maximum since these plots have been made for the
maximum mixing angle.)
This, of course, must be weighted by the neutrino interaction rate,
which is a function of energy if the integral is made over much of a range
of energies.  The global attenuation factor at an energy $E$ is given by:
\[
\frac{1}{2} \times \int_{-1}^{1} \sin^{2}(1.27 \Delta m^{2}
\frac{\sqrt{R_{1}^{2} (\cos^{2}(\theta_{z})-1) + R_{2}^{2}} 
-R_{1} \cos(\theta_{z})}{E}) d \cos(\theta_{z})
\]

\begin{figure}
\psfig{figure=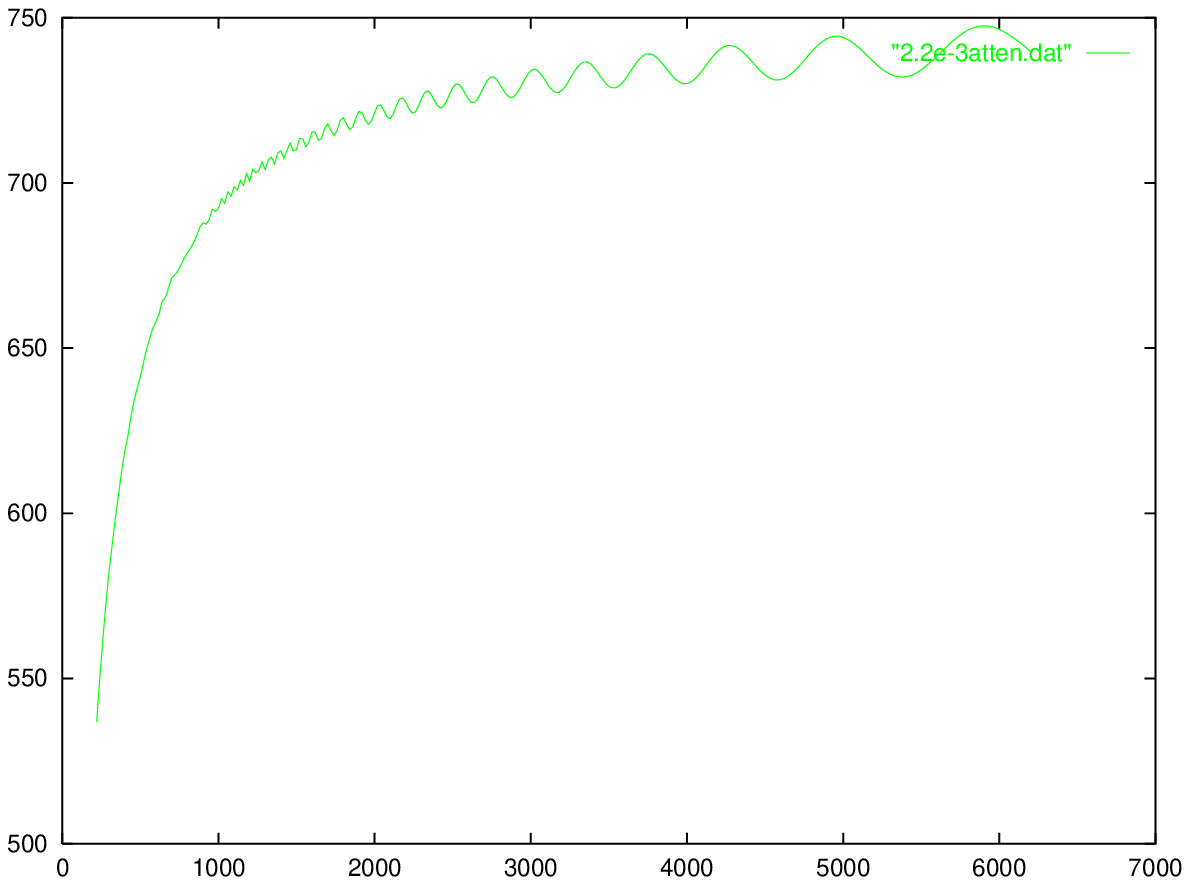,width=5.1in,height=3.0in}
\caption{\label{at2.2} Maximum global attenuation of the neutrino flux as a
function of energy.  A mass
scale $\Delta m^{2}=2.2 \times 10^{-3}$ eV$^{2}$ was used for this plot.}
\end{figure}

\begin{figure}
\psfig{figure=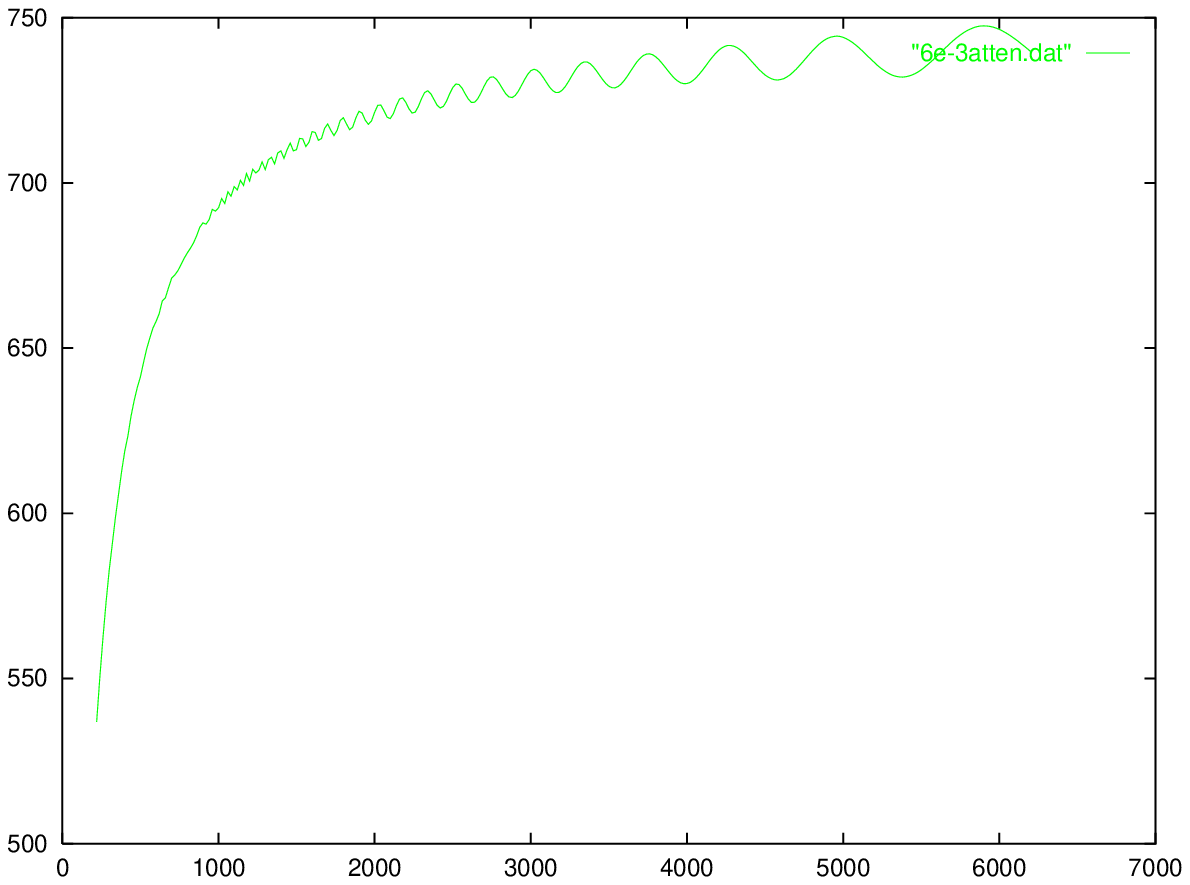,width=5.1in,height=3.0in}
\caption{\label{at6} Maximum global attenuation of the neutrino flux as a
function of energy.  A mass
scale $\Delta m^{2}=6 \times 10^{-3}$ eV$^{2}$ was used for this plot.}
\end{figure}

\begin{figure}
\psfig{figure=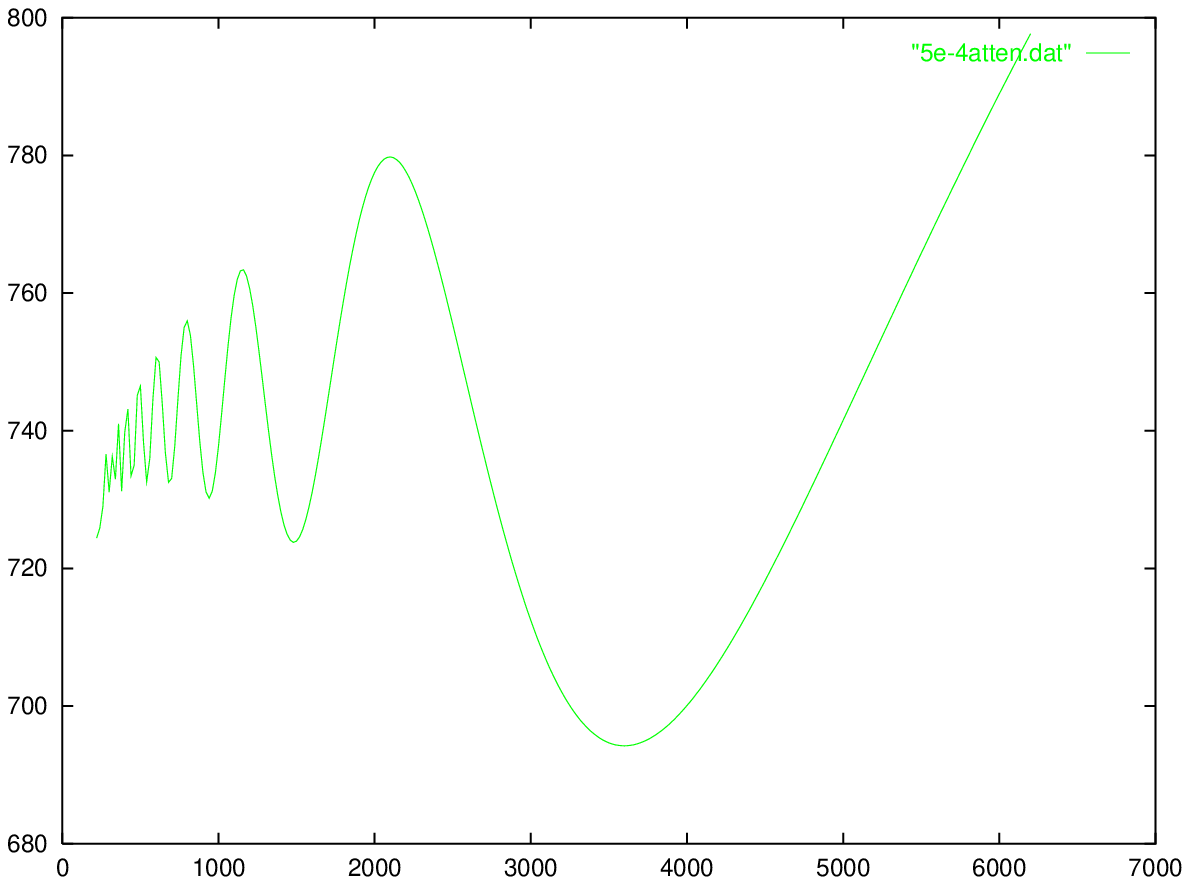,width=5.1in,height=3.0in}
\caption{\label{at5} Maximum global attenuation of the neutrino flux as a
function of energy.  A mass
scale $\Delta m^{2}=5 \times 10^{-4}$ eV$^{2}$ was used for this plot.}
\end{figure}

Figures \ref{at2.2}, \ref{at6} and \ref{at5} are
plots of numerical evaluations of these integrals, as a function of energy.
Figure \ref{at5} shows that at the rather small value of
$\Delta m^{2} = 5 \times 10^{-4}$ eV$^{2}$, the wavelength at modest
energies starts to become comparable to the Earth's diameter.  So at
3.5 GeV a very large portion of the lower hemisphere is maximally
oscillated away.

Figures \ref{at2.2} \ref{at6} and \ref{at5} are the $R$ values that would
be found if the anomaly could be explained by a simple neutrino
oscillation hypothesis that would remove muon neutrinos by turning them
into inactive objects, such as tau neutrinos.  We note that it seems to be
hard to get a value of $R<0.75$ this way.

To be isotropic the integrals over subranges in solid angle should also give the
same value.  For example:
\[
\int_{.6}^{1} \sin^{2}(1.27 \Delta m^{2} \frac{L}{E}) d \cos(\theta_{z})
=\int_{-1}^{-.6} \sin^{2}(1.27 \Delta m^{2} \frac{L}{E}) d \cos(\theta_{z})
\]

It should be clear from figures \ref{2.2e}, \ref{6e} and \ref{5e} that these
integrals will differ.  For all mass scales permitted by the recent
results these integrals differ by about
$\frac{1}{2} \sin^{2}(2 \theta_{Mix})$ since the upper region is not
mixed and the lower region has been, in general, mixed over several cycles.
In cases where the oscillation length is comparable to the Earth's
diameter even larger asymmetries would be seen.  The two integrals are equal
in both the low and high $\Delta m^{2}$ limit.  At very small $\Delta m^{2}$
both integrals are zero.  At very high values of $\Delta m^{2}$ both sides
average over many cycles and are hence $\frac{1}{2}$.

The puzzling feature that the attenuation attributable to oscillations
for these values of $\Delta m^{2}$ is smaller
than that observed could be due to a number
of factors.  If the $\Delta m^{2}$ is large and mixing is maximal as much as
50\% of the muon sample could be removed.  If muon neutrinos oscillate into
electron neutrinos, either via a direct path or via other neutrinos this would
raise the observed electron rate above expectations and therefore lower the
value of $R$.  But the Super-Kamioka data\cite{suposc} shows no evidence for
electron modulation.  The Chooz reactor experiment\cite{chooz} has fairly
strong limits on electron antineutrino oscillations at comparable
$\Delta m^{2}$.  One might evade the Chooz limit with CP violation since Chooz
limits $\bar{\nu_{e}} \rightarrow$ not $\bar{\nu_{e}}$ and one might fix the
atmospheric $R$ with $\nu_{\mu} \rightarrow \nu_{e}$.  But CP violation
effects are unlikely to work in a 3 neutrino scheme since the
$\nu_{\mu} \rightarrow \nu_{e}$ transition must have a short wavelength
and the other amplitudes needed to interfere with it to get CP violation
have at least one long wavelength part.  It might be possible in a 4 neutrino
scheme with 14 parameters.  But the $\bar{\nu_{\mu}}/\nu_{\mu}$ ratio observed
seems to be normal.

A puzzling feature of most atmospheric neutrino experiments has been that the
rate of neutrino interactions agrees fairly well with expectations even though
the number of muon neutrino induced events is low.  
The observed event rate\cite{supcont} is 96$\pm$2\% of the
expected value (94$\pm$3\% of the expected rate for the multi GeV
sample\cite{multig}).
Since the muon neutrino rate appears to be reduced by
about 39\% and the muon neutrino rate is expected to be about twice that
of the electron neutrino rate one might expect an observed flux of only
about 75\% of that expected.  The true neutrino flux may in fact be
25\% higher than predicted.

This might imply that the
electron neutrino rate is too high and compensates for the loss.  It is
also possible that there are sources other than the atmosphere for neutrinos
observed in these detectors.\cite{darkm}


The Super-Kamioka result\cite{suposc} came as a bit of a surprise to the
community since the mass scale in question had been probed and ruled out
by prior experiments.  For example the Super-Kamioka 90\% confidence level
contour of neutrino oscillation parameters is barely consistent with that
from Kamioka.  IMB\cite{clark97} failed to confirm the modulation observed
in the Kamioka Multi GeV data\cite{multigev}.  Aside from concluding that some
of these observations were wrong is there a physical picture which is compatible
with most of them?  The mass of the neutrino is a constant
of nature that can not be different at different times or places.
Since the Earths magnetic field is not a simple dipole aligned along
the Earth's axis one does not expect the cosmic ray flux to be
uniform\cite{honda}.
The most difficult observation to interpret
is the isotropy of the electron sample in \cite{suposc} in the presence of a
large muon anisotropy.  While geomagnetic
effects and the solar cycle can be expected to modulate the atmospheric
spectrum and to give it temporal variation.  These changes would be manifest
in all neutrinos of atmospheric origin.

It may be difficult to rule out {\em any} neutrino oscillation hypothesis
from the atmospheric results alone.  The prior, unmodulated, results are
insensitive to large $\Delta m^{2}$ scales.  It was these relatively large
$\Delta m^{2}$ that provided the motivation for long baseline reactor and
accelerator neutrino experiments.  The magnitude of the atmospheric effect
means that at least some of the mixings to the muon neutrino would be large.

\section*{Acknowledgments}
I would like to thank John Learned, Danka Kielczewska, Wojtek Gajewski and
Maurice Goldhaber
for helpful communications about the Super Kamiokande results.  I am deeply
indebted to Ikaros Bigi for stimulating discussions and for assuring me of
my sanity.  Some of this work was performed at SLAC.

\end{document}